\documentclass[review, fleqn, 12pt]{elsarticle}

 \usepackage{graphics}
\usepackage{graphicx}
\usepackage{epsfig}
\usepackage{subfig}
\usepackage{amssymb}
\usepackage{amsthm}

\newtheorem{theorem}{Theorem}[section]

\newtheorem{remark}{Remark}[section]

\journal{Elsevier}

\begin{document}

\begin{frontmatter}



\title{Soliton, kink and antikink solutions of a 2-component of the Degasperis-Procesi equation}


\author[]{Jiangbo Zhou\corref{cor1}}
\ead{zhoujiangbo@yahoo.cn}

\author[]{Lixin Tian}
\author[]{Xinghua Fan}

\cortext[cor1]{Corresponding author}
\address{Nonlinear Scientific Research Center, Faculty of Science, Jiangsu University,
 Zhenjiang,  Jiangsu 212013, China}
\begin{abstract}
In this paper, we employ the bifurcation theory of planar dynamical
systems to investigate the traveling wave solutions of a 2-component
of the Degasperis-Procesi equation. The expressions for smooth
soliton, kink and antikink solutions are obtained.
\end{abstract}


\end{frontmatter}


\section{Introduction}
\label{} \setcounter{equation}{0}

Since the theory of solitons has very wide applications in fluid
dynamics, nonlinear optics, biochemistry, microbiology, physics and
many other fields, the study of soliton solutions has become one of
the important issues of nonlinear partial differential equations
\cite{1}-\cite{11}.

 In 1999, Degasperis and Procesi \cite{12} derived a nonlinear dispersive
equation
\begin{equation}
\label{eq1.1} u_t - u_{xxt} + 4uu_x = 3u_x u_{xx} + uu_{xxx},
\end{equation}
which is called the Degasperis-Procesi equation. Here $u(t,x)$
represents the fluid velocity at time $t$ in the $x$ direction in
appropriate nondimensional units (or, equivalently the height of the
water's free surface above a flat bottom). The nonlinear convection
term $uu_x$ in Eq.(\ref{eq1.1}) causes the steepening of wave form,
whereas the nonlinear dispersion effect term
$3u_xu_{xx}+uu_{xxx}=(\frac{1}{2}u^2)_{xxx}$ in Eq.(\ref{eq1.1})
makes the wave form spread. Eq.(\ref{eq1.1}) can be regarded as a
model for nonlinear shallow water dynamics \cite{13}.

An important issue regarding the Degasperis-Procesi equation is to
find its traveling wave solutions. Vakhnenko and Parkes \cite{14}
derived periodic and solitary wave solutions of Eq.(\ref{eq1.1}).
Matsuno \cite{15}-\cite{17} obtained multisoliton, $N$-soliton, cusp
and loop soliton solutions of Eq.(\ref{eq1.1}). Lundmark and
Szmigielski \cite{18} investigated multi-peakon solutions of
Eq.(\ref{eq1.1}). Shock wave solutions of Eq.(\ref{eq1.1}) were
obtained in \cite{19}. Lenells \cite{20} classified all its  weak
traveling wave solutions. Chen and Tang \cite{21}  showed that
Eq.(\ref{eq1.1}) has kink-like and antikink-like wave solutions.
Qiao \cite{8} obtained the peakons, dehisced solitons, cuspons and
new 1-peak solitons of Eq.(\ref{eq1.1}).

It is known that the Degasperis-Procesi equation has solitons, but
has no kink or antikink solutions. In this paper, we generalize the
Degasperis-Procesi equation to the following 2-component of
Degasperis-Procesi equation
\begin{equation}
\label{eq1.2} \left\{ {\begin{array}{l}
m_t+u m_x+3u_x m =v_x, \\
 v_t=(uv)_x, \\
 \end{array}} \right.
\end{equation}
where $m=u-u_{xx}$. This 2-component of the Degasperis-Procesi
equation can be applied to describe shallow water waves. In
Eqs.(\ref{eq1.2}),  $u(x, t)$ denotes the height of the water
surface above a horizontal bottom and $v(x, t)$ is related to the
horizontal velocity field. Obviously, if $v\equiv 0$, then
Eqs.(\ref{eq1.2}) is reduced to the Degasperis-Procesi equation
(\ref{eq1.1}). Using the bifurcation theory of planar dynamical
systems (see \cite {22}), we will show that, in addition to smooth
solitons, Eqs.(\ref{eq1.2}) has kink and antikink solutions. The
solitons denote the nonlinear localized waves on the shallow water's
free surface that retain their individuality under interaction and
eventually travel with their original shapes and speeds. The kink
and antikink waves, which rise or descend from one asymptotic state
to another, form another type. The balance between the nonlinear
steepening and dispersion effect under Eqs.(\ref{eq1.2}) gives rise
to these solutions. The solutions presented in this paper may help
people to know deeply the described physical process of
Eqs.(\ref{eq1.2}).

The remainder of the paper is organized as follows. In Section 2,
using the traveling wave transformation,  we transform
Eqs.(\ref{eq1.2}) into a planar dynamical system and then discuss
bifurcations of phase portraits of this system. In Section 3, we
obtain the expressions for the smooth soliton, kink and antikink
solutions of Eqs.(\ref{eq1.2}). A short conclusion is given in
Section 4.

\section{Bifurcation and phase portraits of the traveling wave system}
 \label{}
 \setcounter{equation}{0}
Eqs.(\ref{eq1.2}) also takes the form
\begin{equation}
\label{eq2.1} \left\{ {\begin{array}{l}
u_t - u_{xxt} + 4uu_x = 3u_x u_{xx} + uu_{xxx} +v_x, \\
 v_t=(uv)_x .\\
 \end{array}} \right.
\end{equation}

Let $\xi = x+ct$, where $c$ is the wave speed. By using the
traveling wave transformation $u(x, t) =\varphi (x+ct)= \varphi (\xi
)$, $v(x, t) =\psi (x+ct)= \psi (\xi )$, we reduce system
(\ref{eq1.2}) to the following ordinary differential equations:
\begin{equation} \label{eq2.2} \left\{ {\begin{array}{l}
c\varphi' - c\varphi''' +4\varphi \varphi' = 3\varphi' \varphi'' + \varphi \varphi''' +\psi' ,\\
 c\psi'=(\varphi\psi)' .\\
 \end{array}} \right.
\end{equation}

Integrating (\ref{eq2.2}) once with respect to $\xi$, we have
\begin{equation}
\label{eq2.3} \left\{ {\begin{array}{l}
c\varphi - c\varphi'' =2 \varphi^2-(\varphi')^2- \varphi \varphi'' +\psi+g, \\
 c\psi=\varphi\psi+h, \\
 \end{array}} \right.
\end{equation}
\noindent where $g, h$ are two integral constants.

From the second expression in system (\ref{eq2.3}), we can obtain
that
\begin{equation}
\label{eq2.4} \psi=-\frac{h}{\varphi-c}.
\end{equation}

Substituting (\ref{eq2.4}) into the first expression in system
(\ref{eq2.3}) yields

\begin{equation}
\label{eq2.5}  \varphi'' =\frac{(2 \varphi^2-c \varphi-(\varphi')^2
+g)(\varphi - c)-h }{(\varphi - c)^2}.
\end{equation}

Let $y = \varphi '$, then we get the following planar dynamical
system:
\begin{equation}
\label{eq2.6}\left\{ {\begin{array}{l}
 \frac{\textstyle  d\varphi }{\textstyle  d\xi } = y, \\
 \frac{\textstyle dy}{\textstyle d\xi } = \frac{\textstyle(2 \varphi^2-c \varphi-y^2
+g)(\varphi - c)-h }{\textstyle(\varphi - c)^2}. \\
 \end{array}} \right.
\end{equation}
\noindent This system has a first integral
\begin{equation}
\label{eq2.7} H(\varphi,
y)=\varphi^4-2c\varphi^3+(c^2+g)\varphi^2-2(cg+h)\varphi-y^2(\varphi-c)^2=k,
\end{equation}
\noindent where $k$ is a constant.

Note that (\ref{eq2.6}) has a singular line $\varphi = c$. To avoid
the line temporarily, we make transformation $d\xi = (\varphi - c)^2
d\zeta $. Under this transformation, Eq.(\ref{eq2.6}) becomes
\begin{equation}
\label{eq2.8} \left\{ {\begin{array}{l}
 \frac{\textstyle d\varphi }{\textstyle d\zeta } = y(\varphi - c)^2, \\
 \frac{\textstyle dy}{\textstyle d\zeta } =( 2 \varphi^2-c \varphi-y^2
+g)(\varphi - c)-h.
\\
 \end{array}} \right.
\end{equation}

System (\ref{eq2.6}) and system (\ref{eq2.8}) have the same first
integral as (\ref{eq2.7}). Consequently, system (\ref{eq2.8}) has
the same topological phase portraits as system (\ref{eq2.6}) except
for the straight line $\varphi = c$.

For a fixed $k$, (\ref{eq2.7}) determines a set of invariant curves
of system (\ref{eq2.8}). As $k$ is varied, (\ref{eq2.7}) determines
different families of orbits of system (\ref{eq2.8}) having
different dynamical behaviors. Let $M(\varphi _e , y_e )$ be the
coefficient matrix of the linearized version of system (\ref{eq2.8})
at the equilibrium point $(\varphi _e , y_e )$, then
\begin{equation}
\label{eq2.9} M(\varphi _e ,y_e ) = \left( {{\begin{array}{*{20}c}
{\indent  \indent  2 y_e (\varphi _e-c) } \hfill &&& {\indent \indent (\varphi _e - c)^2} \hfill \\
 {6\varphi _e^2 -6c \varphi _e+c^2+g-y^2 } \hfill &&& \indent{- 2y_e(\varphi _e-c)} \hfill \\
\end{array} }} \right)
\end{equation}
and at this equilibrium point, we have
\begin{equation}
\label{eq2.10} J(\varphi _e ,y_e ) = \det M(\varphi _e ,y_e ) = -
3y_e^2(\varphi _e-c)^2 - (6\varphi _e^2 -6c \varphi
_e+c^2+g)(\varphi _e - c)^2,
\end{equation}
\begin{equation}
\label{eq2.11} p(\varphi _e ,y_e ) = \mathrm{trace}(M(\varphi _e
,y_e )) = 0.
\end{equation}
By the qualitative theory of differential equations (see \cite
{22}), for an equilibrium point of a planar dynamical system, if $J
< 0$, then this equilibrium point is a saddle point; it is a center
point if $J
> 0$ and $p = 0$; if $J = 0$ and the Poincar\'{e} index of the
equilibrium point is 0, then it is a cusp.

\begin{figure}[h]
\centering \subfloat[]
{\includegraphics[height=1.2in,width=1.4in]{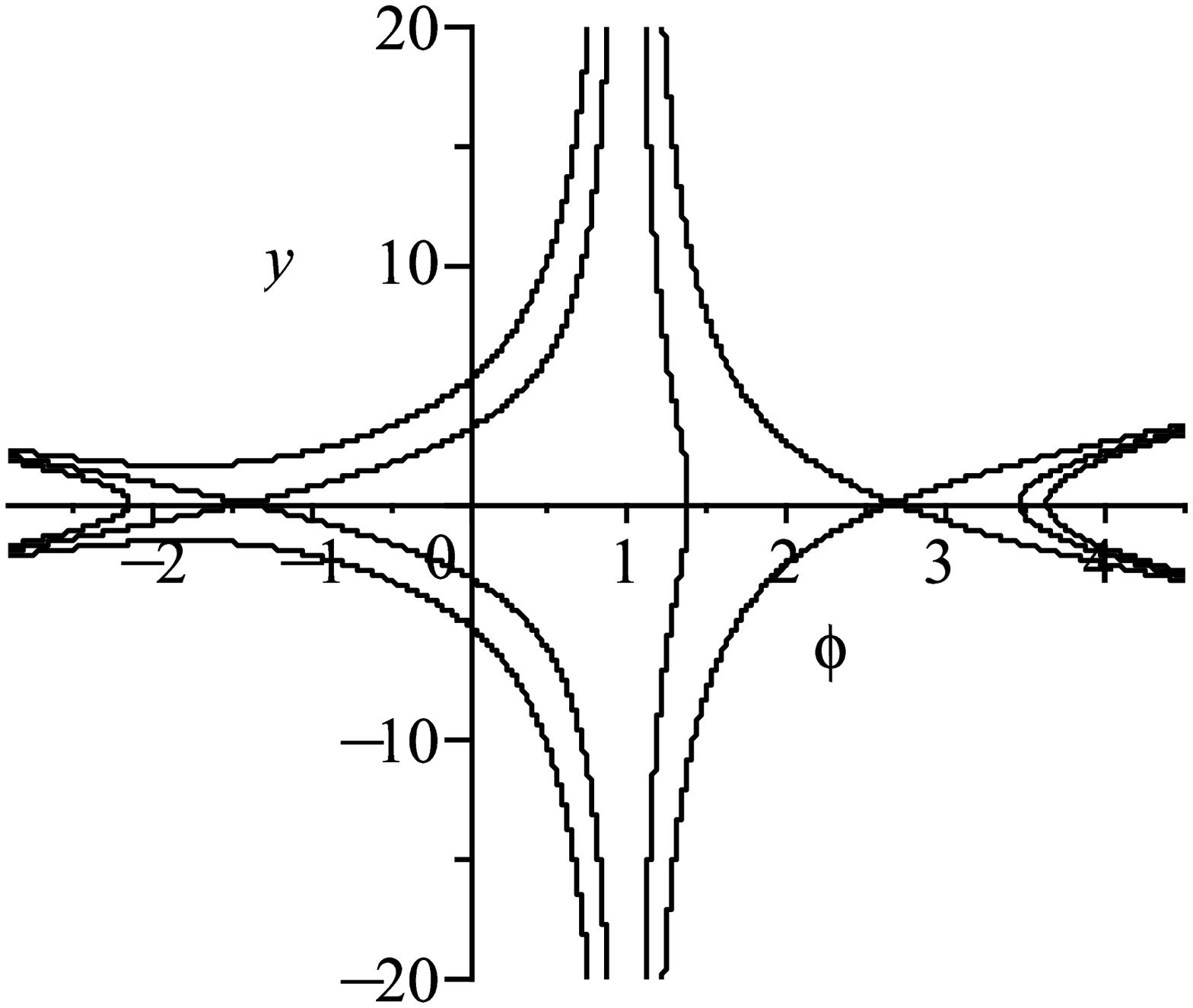}}
\subfloat[]{\includegraphics[height=1.2in,width=1.4in]{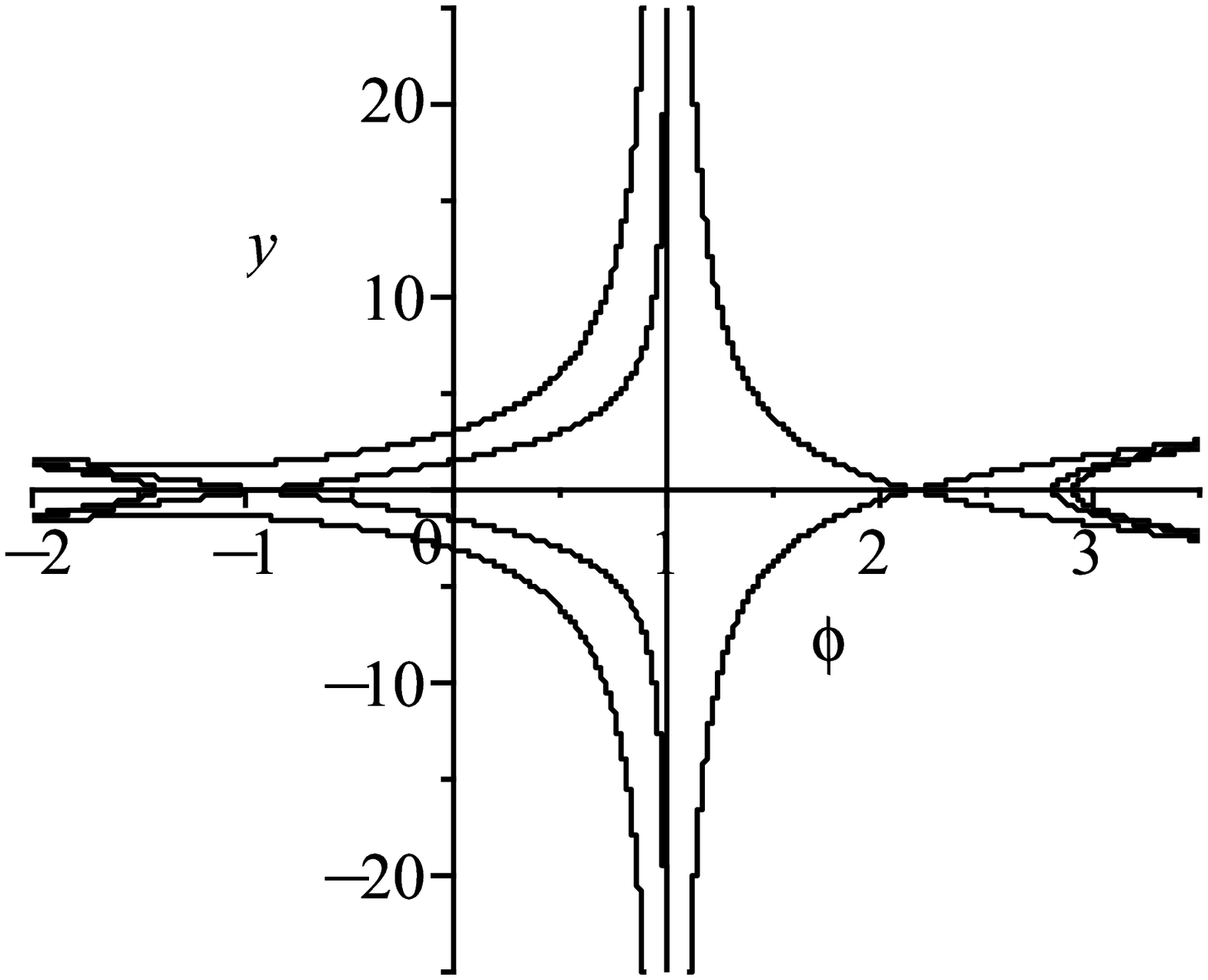}}
\subfloat[]{\includegraphics[height=1.2in,width=1.4in]{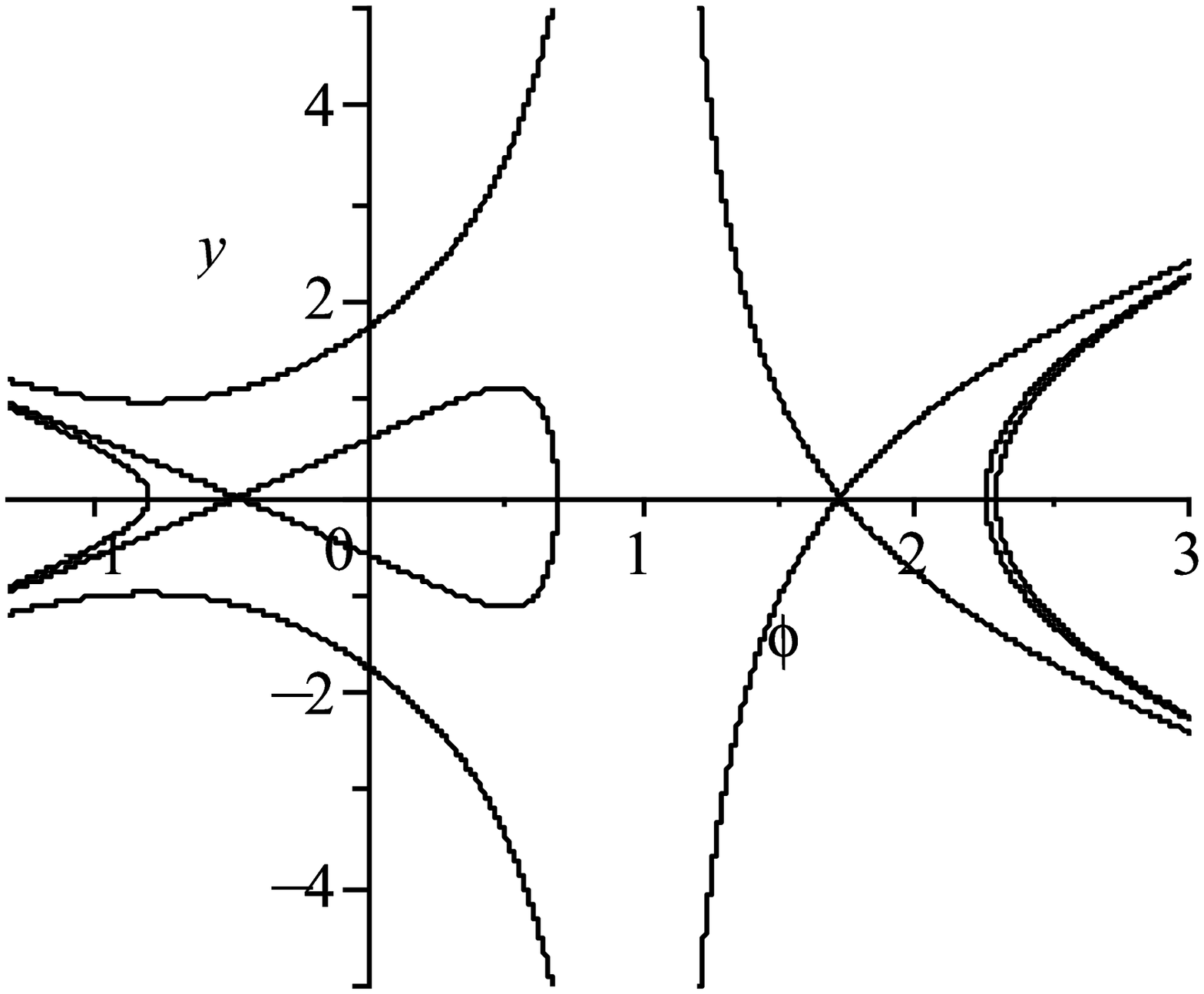}}
\subfloat[]{\includegraphics[height=1.2in,width=1.4in]{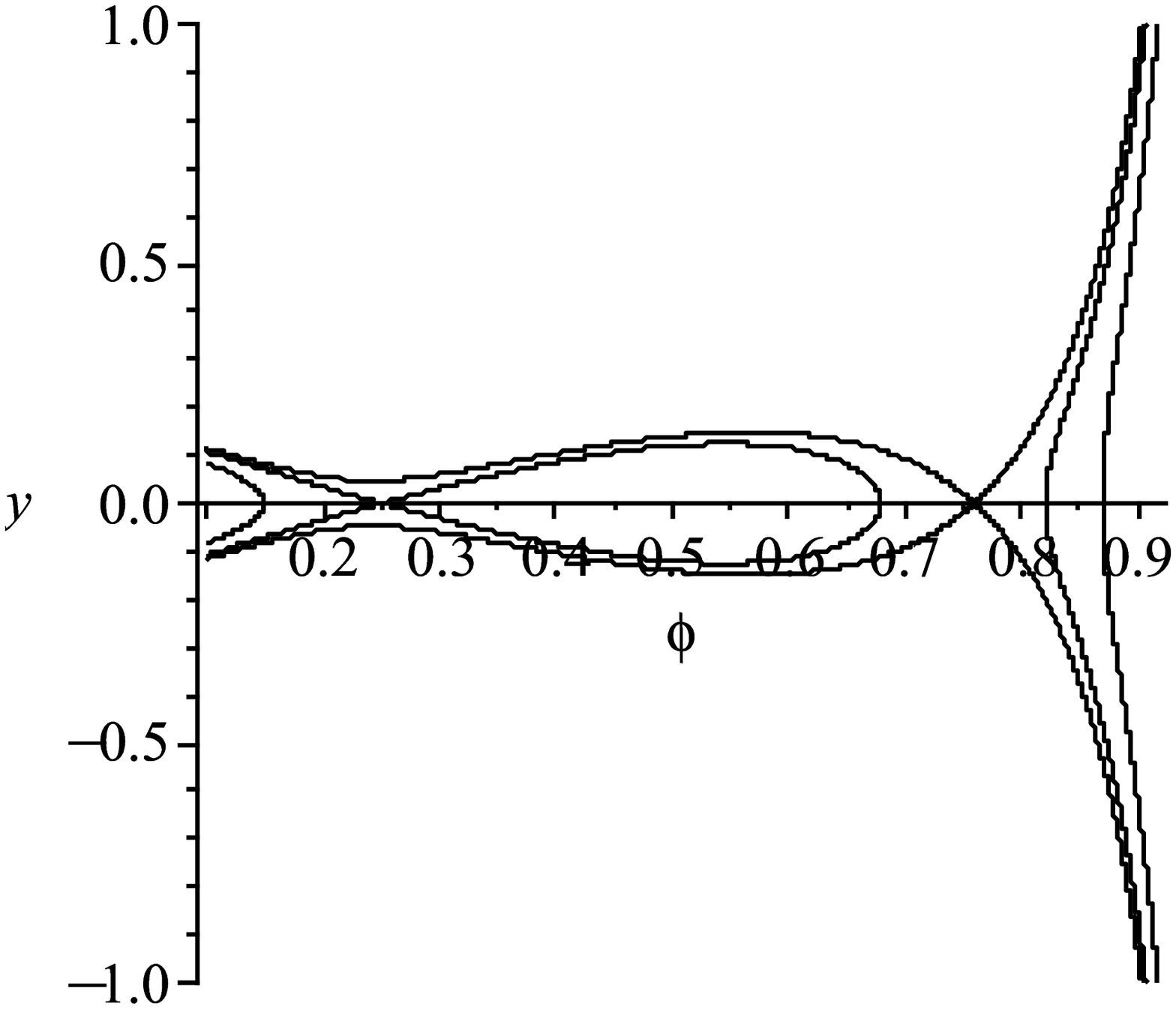}}\\
\subfloat[]{\includegraphics[height=1.2in,width=1.4in]{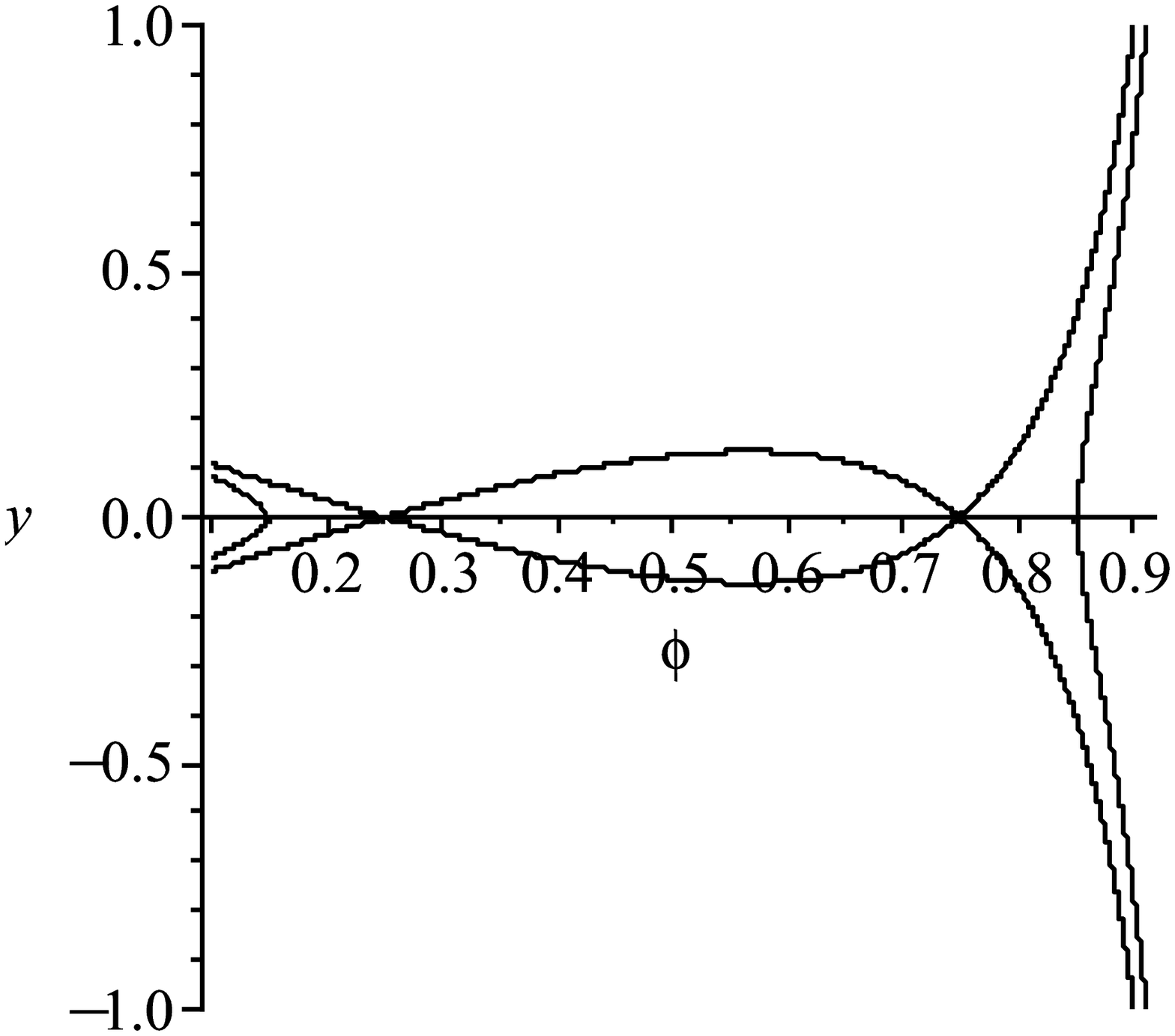}}
\subfloat[]{\includegraphics[height=1.2in,width=1.4in]{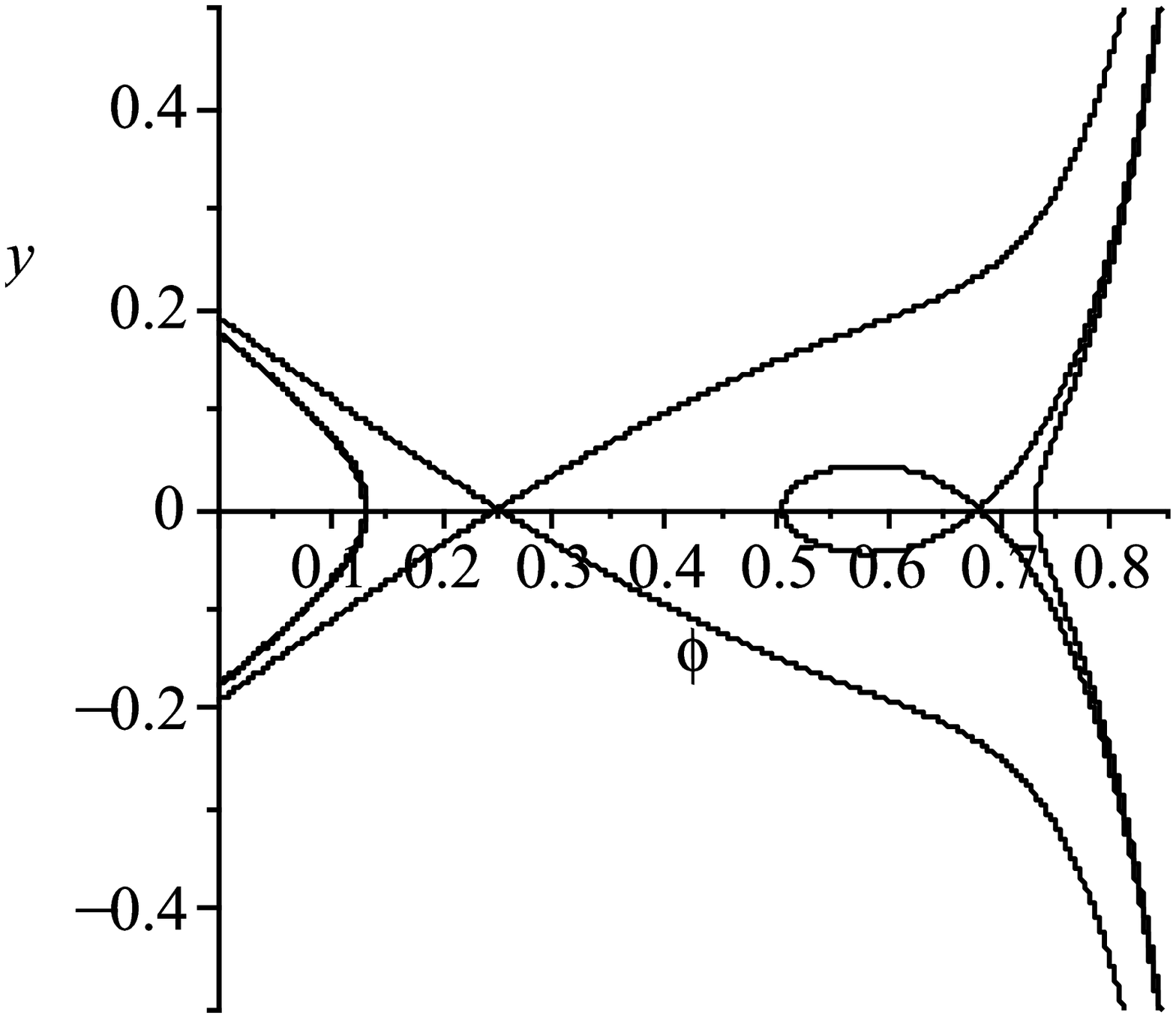}}
\subfloat[]{\includegraphics[height=1.2in,width=1.40in]{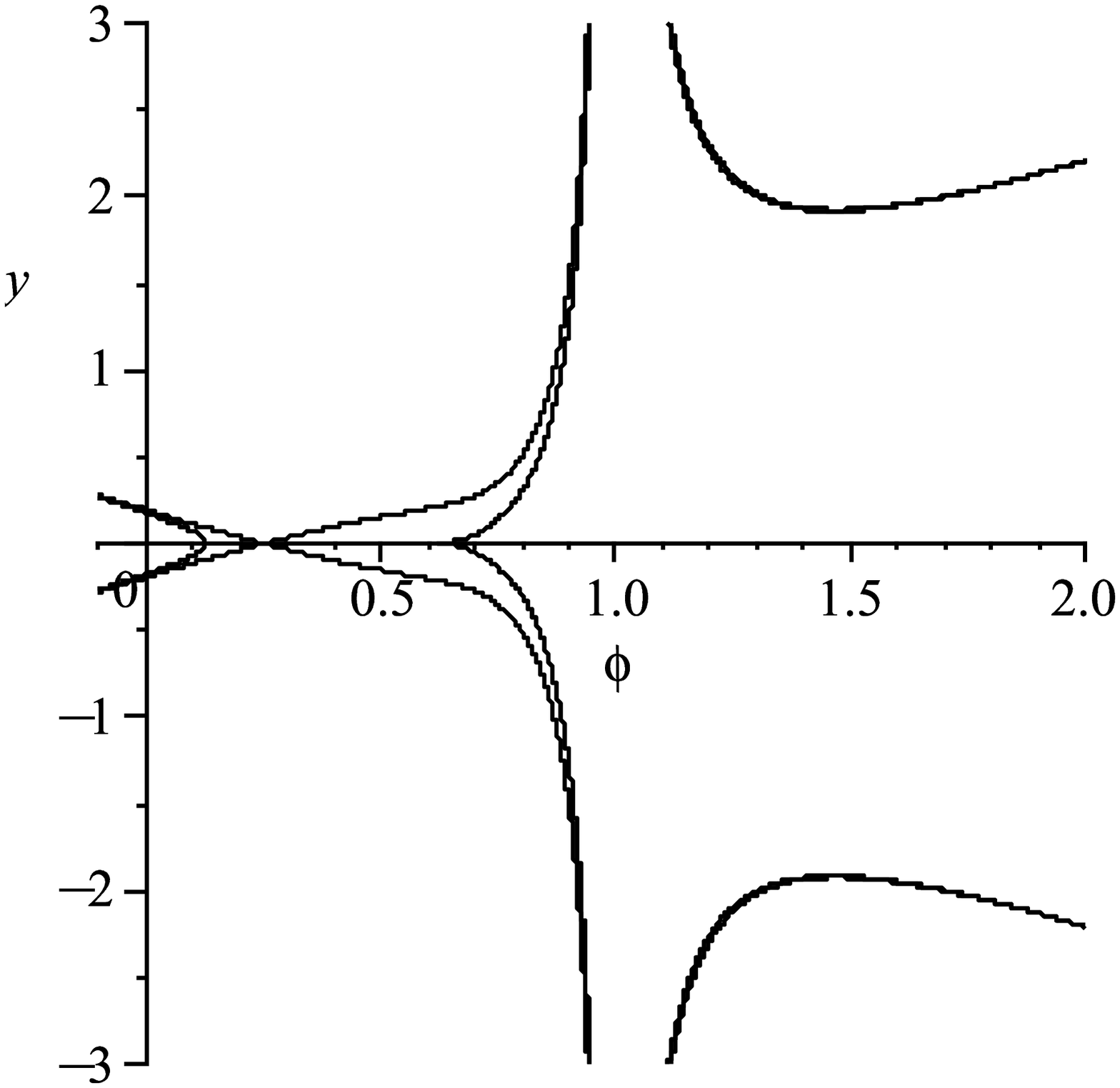}}
\subfloat[]{\includegraphics[height=1.2in,width=1.4in]{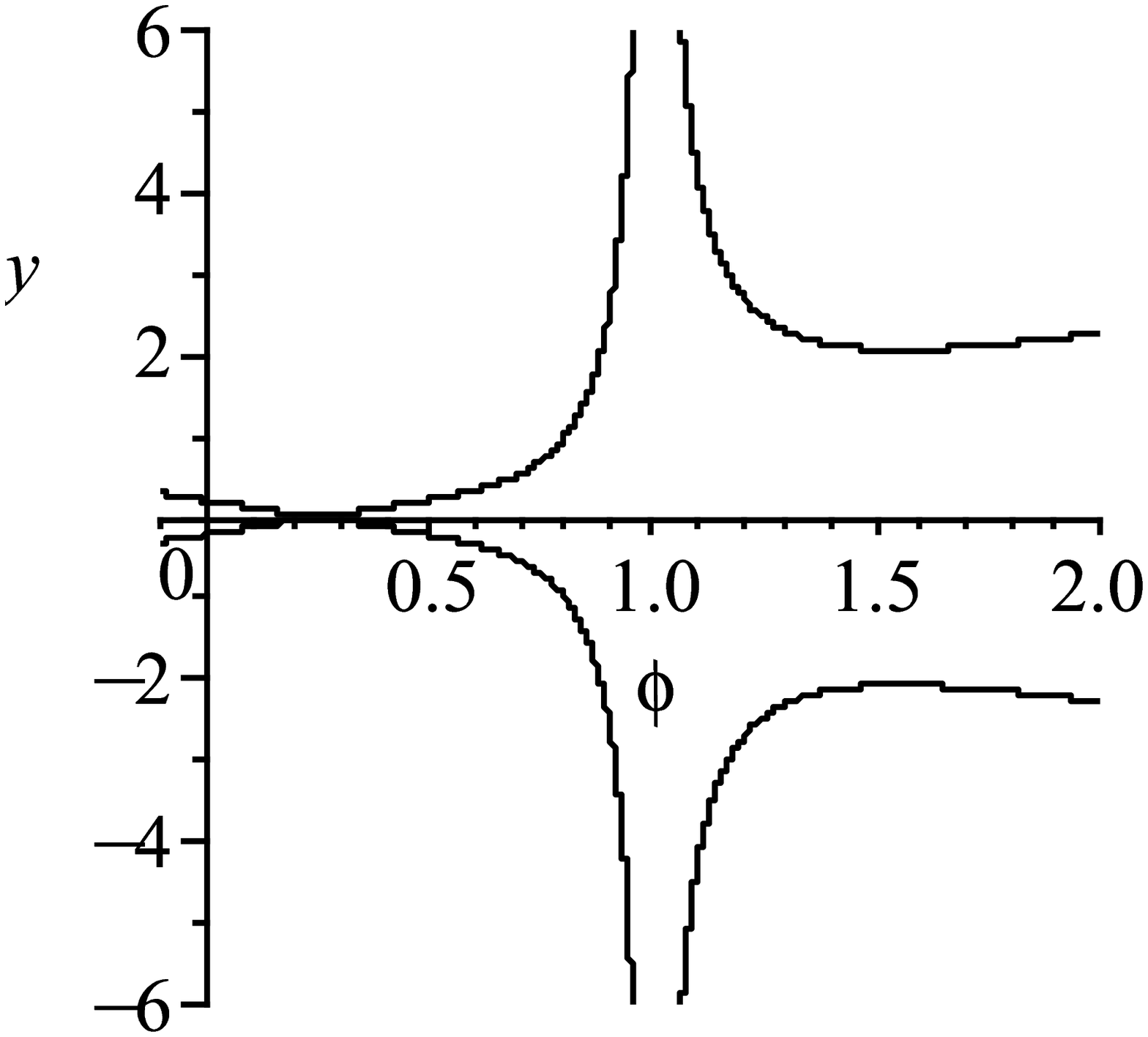}}
\caption{The phase portraits of system (\ref{eq2.8}) for the
parameter $c>0$. (a) $g<g_1(c)$; (b) $g=g_1(c)$; (c)
$g_1(c)<g<g_2(c)$; (d) $g_2(c)<g<g_3(c)$; (e)$g=g_3(c)$; (f)
$g_3(c)<g<g_4(c)$; (g) $g=g_4(c)$; (h) $g>g_4(c)$. }\label{fig1}
\end{figure}

Noting that it is impossible to figure out the equilibrium points of
system (\ref{eq2.8}) for arbitrary constants $c$, $g$ and $h$, so
here for the sake of convenience, we set
$h=\frac{3c}{32}(c^2-8g)\neq 0$. By using the first integral value
and properties of equilibrium points, we obtain the bifurcation
curves as follows:
\begin{equation}
\label{eq2.12} g_1 (c) =-\frac{(16+7\sqrt{7})c^2}{8},
\end{equation}
\begin{equation}
\label{eq2.13} g_2 (c) = \frac{c^2}{8},
\end{equation}
\begin{equation}
\label{eq2.14} g_3 (c) =  \frac{3c^2}{8},
\end{equation}
\begin{equation}
\label{eq2.15} g_4 (c) = \frac{13c^2}{32}.
\end{equation}
Obviously, the above four curves have no intersection point and
$g_1(c)<g_2(c)<g_3(c)<g_4(c)$ for arbitrary constant $c\neq 0$.

Using bifurcation method of vector fields (e.g., \cite{20}), we have
the following result which describes the locations and properties of
the equilibrium points of system (\ref{eq2.8}).

\begin{theorem}
For given any constant wave speed $c\neq 0$, let
\begin{equation}
\label{eq2.16} \varphi _{1\pm} = \frac{1}{8}(5c\pm \sqrt {13 c^2 -
32g})\quad for \quad g\leq g_4(c).
\end{equation}
Then we have:

(1) If $g<g_2(c)$, then system (\ref{eq2.8}) has three equilibrium
points $(\varphi _{1 -} ,0)$ , $(\varphi _{1+} ,0)$ and $(
\frac{c}{4} , 0)$.  $(\varphi _{1 -}, 0)$ and $(\varphi _{1+}, 0)$
are two saddle points, while $( \frac{c}{4} , 0)$ is a center point.
Specially, when $g_1(c)< g<g_2(c)$, system (\ref{eq2.8}) has a
homoclinic orbit, which connects with the saddle point $(\varphi _{1
-} ,0)$.

(2) If $g_2(c)<g<g_4(c)$, then system (\ref{eq2.8}) has three
equilibrium points $(\varphi _{1 -} , 0)$ , $(\varphi _{1+} , 0)$
and $( \frac{c}{4} , 0)$.   $( \frac{c}{4} , 0)$ and $(\varphi
_{1+}, 0)$ are two saddle points, while $(\varphi _{1 -}, 0)$ is a
center point. Specially,

(i) When $g_2(c)<g<g_3(c)$, system (\ref{eq2.8}) has a homoclinic
orbit, which connects with the saddle point $( \frac{c}{4}, 0)$.

(ii) When $g=g_3(c)$,  system (\ref{eq2.8}) has two heteroclinic
orbits. They connect with the saddle points $( \frac{c}{4}, 0)$ and
$( \frac{3c}{4}, 0)$.

(iii) When $g_3(c)<g<g_4(c)$, system (\ref{eq2.8}) has a homoclinic
orbit, which connects with the saddle point $(\varphi _{1+}, 0)$.

(3)  If  $g=g_4(c)$, then system (\ref{eq2.8}) has two equilibrium
points $( \frac{c}{4}, 0)$ and  $(\frac{5c}{8}, 0)$. $( \frac{c}{4},
0)$ is a saddle point, while $( \frac{5c}{8} , 0)$ is a cusp.

(4) If  $g>g_4(c)$, then system (\ref{eq2.8}) has only one
equilibrium point $( \frac{c}{4}, 0)$. It is a saddle point.
\end{theorem}

Without loss of generality, we show the phase portraits of system
(\ref{eq2.8}) in each region and on the bifurcation curves in
Fig.\ref{fig1} for the wave speed $c>0$.

\section{Soliton, kink and antikink solutions of Eqs.(\ref{eq1.2})}
 \label{}
 \setcounter{equation}{0}
Suppose that $\varphi(\xi )(\xi = x + ct)$ is a traveling wave
solution of the first component $u$ in Eqs.(\ref{eq1.2}) for $\xi\in
( - \infty, + \infty )$, and $\mathop {\lim }\limits_{\xi \to -
\infty } \varphi(\xi ) = A$, $\mathop {\lim }\limits_{\xi\to+\infty
} \varphi(\xi ) = B$, where $A$ and $B$ are two constants. If $A=B$,
then $\varphi(\xi )$ is called a soliton solution. If $A \ne B$,
then $\varphi(\xi )$ is called a kink (or an antikink) solution.
Usually, a soliton solution of Eqs.(\ref{eq1.2}) corresponds to a
homoclinic orbit of system (\ref{eq2.6}) and  a kink (or an
antikink) solution of Eq.(\ref{eq1.1}) corresponds to a heteroclinic
orbit (or the so-called connecting orbit) of  system (\ref{eq2.6}).

In Fig.\ref{fig1}(c), (d), (f), the homoclinic orbits of system
(\ref{eq2.6}) or (\ref{eq2.8}) can be expressed respectively as
\begin{equation}
\label{eq3.1}  y = \pm\frac{( \varphi-\varphi _{1-})\sqrt {\varphi
^2 + l_{1-} \varphi + l_{2-} } }{\varphi-c} \quad \textrm{for} \quad
\varphi _{1-}\leq\varphi \leq\varphi _{2-},
\end{equation}
\begin{equation}
\label{eq3.2}  y = \pm\frac{( \varphi-\frac{c}{4})\sqrt {\varphi ^2
+ l_1 \varphi + l_2 } }{\varphi-c} \quad \textrm{for} \quad
\frac{c}{4}\leq\varphi \leq\varphi _1,
\end{equation}
\begin{equation}
\label{eq3.3}  y = \pm\frac{( \varphi-\varphi _{1+})\sqrt {\varphi
^2 + l_{1+} \varphi + l_{2-} } }{\varphi-c} \quad \textrm{for} \quad
\varphi _{2+}\leq\varphi \leq\varphi _{1+},
\end{equation}
where
\begin{equation}
\label{eq3.4} l_{1\pm} = \frac{1}{4}(- 3c \pm \sqrt {13c^2 - 32g} ),
\end{equation}
\begin{equation}
\label{eq3.5}l_{2\pm} = -\frac{1}{32}(-9c^2 +16g \pm c \sqrt {13c^2
- 32g}),
\end{equation}
\begin{equation}
\label{eq3.6} l_1 = -\frac{3c}{2},
\end{equation}
\begin{equation}
\label{eq3.7}l_2 = -\frac{3}{16}c^2 g ,
\end{equation}
\begin{equation}
\label{eq3.8}\varphi_1=\frac{1}{4}(3c-\sqrt {6c^2 - 16g}),
\end{equation}
\begin{equation}
\label{eq3.9}\varphi_{2\pm}=\frac{1}{8}(3c\pm(-\sqrt {13c^2 -
32g}+2\sqrt {c(c+\sqrt {13c^2 - 32g})})),
\end{equation}
\noindent and $\varphi_{1\pm}$ is as in (\ref{eq2.16}).

Substituting Eq.(\ref{eq3.1})-(\ref{eq3.3}) into the first equation
of system (\ref{eq2.6}), respectively, and integrating along the
corresponding homoclinic orbit, we have
\begin{equation}
\label{eq3.10} {\int_\varphi^{\varphi _{2-}} {\frac{s-c}{(s -
\varphi _{1-} )\sqrt {s^2 + l_{1-} s + l_{2-} } }ds = -|\xi| }},
\end{equation}
\begin{equation}
\label{eq3.11} {\int_\varphi^{\varphi _1} {\frac{s-c}{(s -
\frac{c}{4} )\sqrt {s^2 + l_1 s + l_2 } }ds = -|\xi| }},
\end{equation}
\begin{equation}
\label{eq3.12} {\int_{\varphi _{2+}}^\varphi {\frac{s-c}{(s -
\varphi _{1+} )\sqrt {s^2 + l_{1+} s + l_{2+} } }ds = -|\xi| }}.
\end{equation}
It follows from (\ref{eq3.10})-(\ref{eq3.12}) that
\begin{equation}
\label{eq3.13} \beta_-(\varphi _{2-}) =
\beta_-(\varphi)\exp(-|\xi|),
\end{equation}
\begin{equation}
\label{eq3.14} \beta(\varphi _1) = \beta(\varphi)\exp(-|\xi|),
\end{equation}
\begin{equation}
\label{eq3.15} \beta_+(\varphi) =\beta_+(\varphi _{2+})
\exp(-|\xi|),
\end{equation}
where
\begin{equation}
\label{eq3.16}\beta_\pm(\varphi) =  \frac{(2\sqrt {\varphi^2 +
l_{1\pm} \varphi + l_{2\pm} } + 2\varphi + l_{1\pm} )(\varphi -
\varphi _{1\pm})^{\alpha _{1\pm} }}{(2\sqrt {a_{1\pm} } \sqrt
{\varphi^2 + l_{1\pm} \varphi + l_{2\pm} } + b_{1\pm} \varphi +
l_{3\pm} )^{\alpha _{1\pm} }},
\end{equation}
\begin{equation}
\label{eq3.17}
 \beta(\varphi) =  \frac{(2\sqrt {\varphi^2 + l_1
\varphi + l_2 } + 2\varphi + l_1 )(\varphi - \frac{c}{4})^{\alpha
_1}} {(2\sqrt {a_1} \sqrt {\varphi^2 + l_1 \varphi + l_2 } + b_1
\varphi + l_3 )^{\alpha _1}},
\end{equation}
\begin{equation}
\label{eq3.18} l_{3\pm} = \frac{1}{2}c^2 - 2g,
\end{equation}
\begin{equation}
\label{eq3.19} l_3 = 2g,
\end{equation}
\begin{equation}
\label{eq3.20} a_{1\pm} = \frac{1}{16}(13c^2 - 32g \pm 3c\sqrt
{13c^2 - 32g}),
\end{equation}
\begin{equation}
\label{eq3.21} a_1 = -\frac{1}{8}c^2 +g,
\end{equation}
\begin{equation}
\label{eq3.22}b_{1\pm} = \frac{1}{2}(c \pm \sqrt {13c^2 - 32g}),
\end{equation}
\begin{equation}
\label{eq3.23}b_1 = -c,
\end{equation}
\begin{equation}
\label{eq3.24} \alpha_{1\pm} = \frac{5c \pm \sqrt {13c^2 - 32g}}{2
\sqrt {13c^2 - 32g - 3c \sqrt {13c^2 - 32g }}},
\end{equation}
\begin{equation}
\label{eq3.25} \alpha_1 = \frac{c}{\sqrt {-2c^2 +16g}},
\end{equation}
and $l_{1\pm}$, $l_{2\pm}$, $l_1 $, $l_2$, $\varphi _{1\pm}$ and
$\varphi _{2\pm}$ are as in (\ref{eq3.4}), (\ref{eq3.5}),
(\ref{eq3.6}), (\ref{eq3.7}), (\ref{eq2.16}) and (\ref{eq3.9}),
respectively.

 (\ref{eq3.13})-(\ref{eq3.15}) are the implicit
expressions for the soliton solutions of the first component $u$ in
Eqs.(\ref{eq1.2}). From (\ref{eq2.4}),
(\ref{eq3.13})-(\ref{eq3.15}), we can give the parametric
expressions for the soliton solutions of the second component $v$ in
Eqs.(\ref{eq1.2}) as follows:
\begin{equation}
\label{eq3.26} \left\{ {\begin{array}{l}
  \xi=\pm (\ln\beta_-(\varphi)-\ln\beta_-(\varphi _{2-})), \\
 \psi=-\frac{\textstyle 3c(c^2-8g)}{\textstyle 32(\varphi-c)},
\\
 \end{array}} \right.      \;\quad ( \varphi _{1-}\leq\varphi \leq\varphi
 _{2-}),
\end{equation}
\begin{equation}
\label{eq3.27} \left\{ {\begin{array}{l}
  \xi=\pm (\ln\beta(\varphi)-\ln\beta_-(\varphi _1)), \\
 \psi=-\frac{\textstyle 3c(c^2-8g)}{\textstyle 32(\varphi-c)},
\\
 \end{array}} \right.\;\quad ( \frac{c}{4}\leq\varphi \leq\varphi
 _1),
\end{equation}
\begin{equation}
\label{eq3.28} \left\{ {\begin{array}{l}
  \xi=\pm (\ln\beta_+(\varphi_{2+})-\ln\beta_+(\varphi)), \\
 \psi=-\frac{\textstyle 3c(c^2-8g)}{\textstyle 32(\varphi-c)},
\\
 \end{array}} \right.\;\quad  (\varphi _{2+}\leq\varphi \leq\varphi
 _{1+}),
\end{equation}
with $\varphi$ as the parameter.

\begin{figure}[h]
\centering
\subfloat[]{\includegraphics[height=1.5in,width=2.4in]{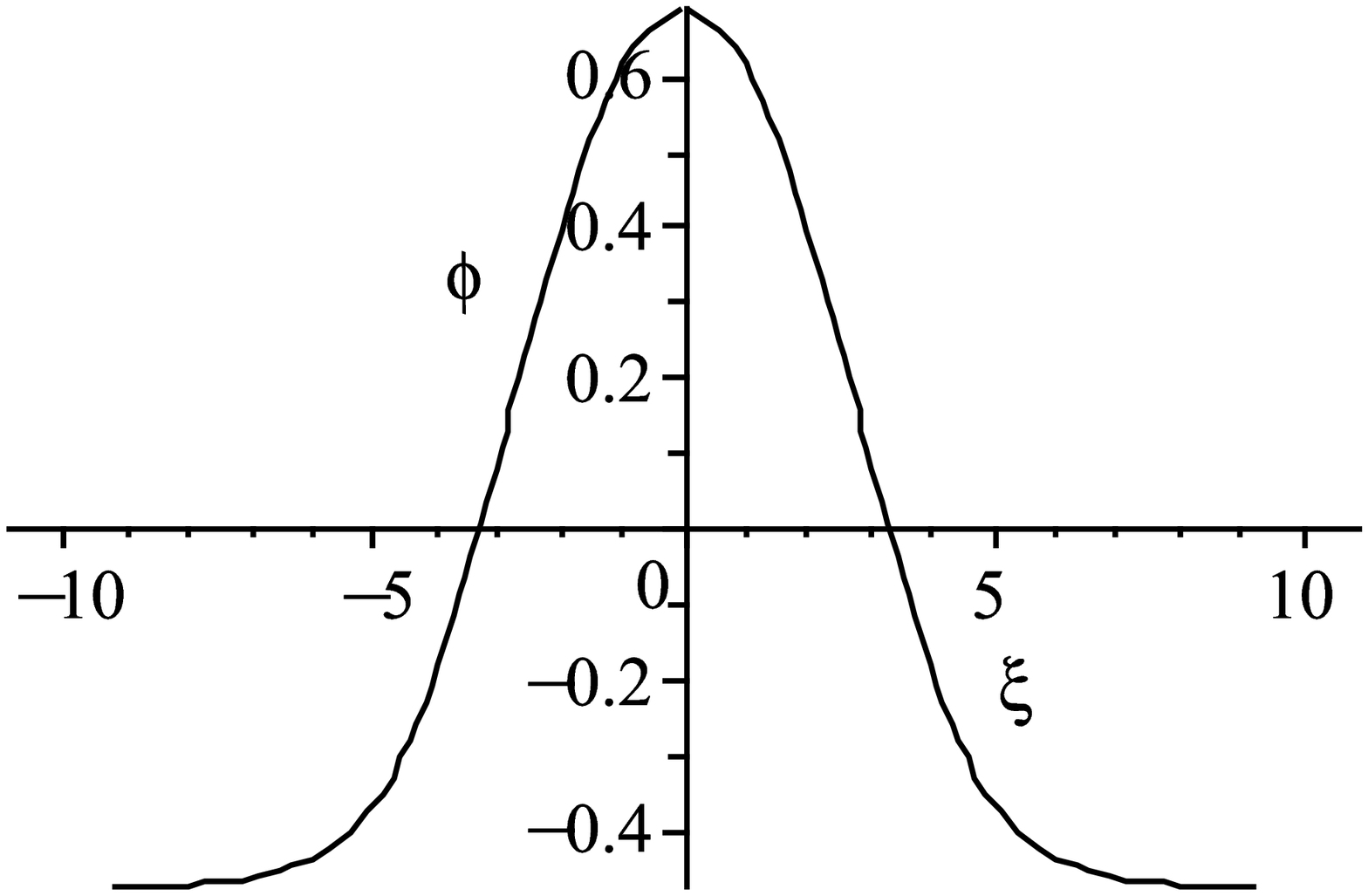}}\hspace{0.08\linewidth}
\subfloat[]{\includegraphics[height=1.5in,width=2.4in]{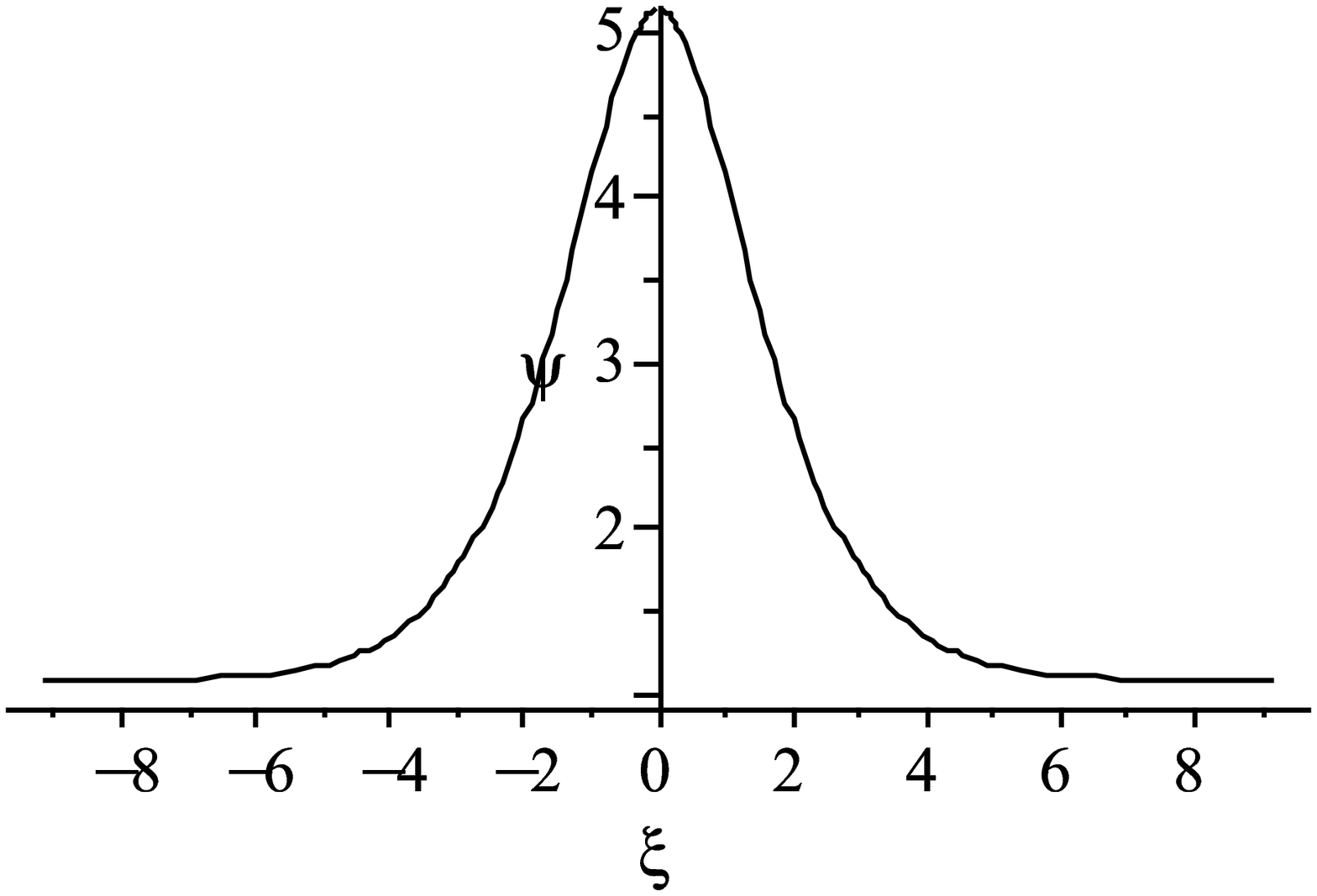}}\\
\caption{The soliton of Eqs.(\ref{eq1.2}). ($c=1$, $g=-2$,
$h=1.593750$)}\label{fig2}
\end{figure}

\begin{figure}[h]
\centering
\subfloat[]{\includegraphics[height=1.5in,width=2.4in]{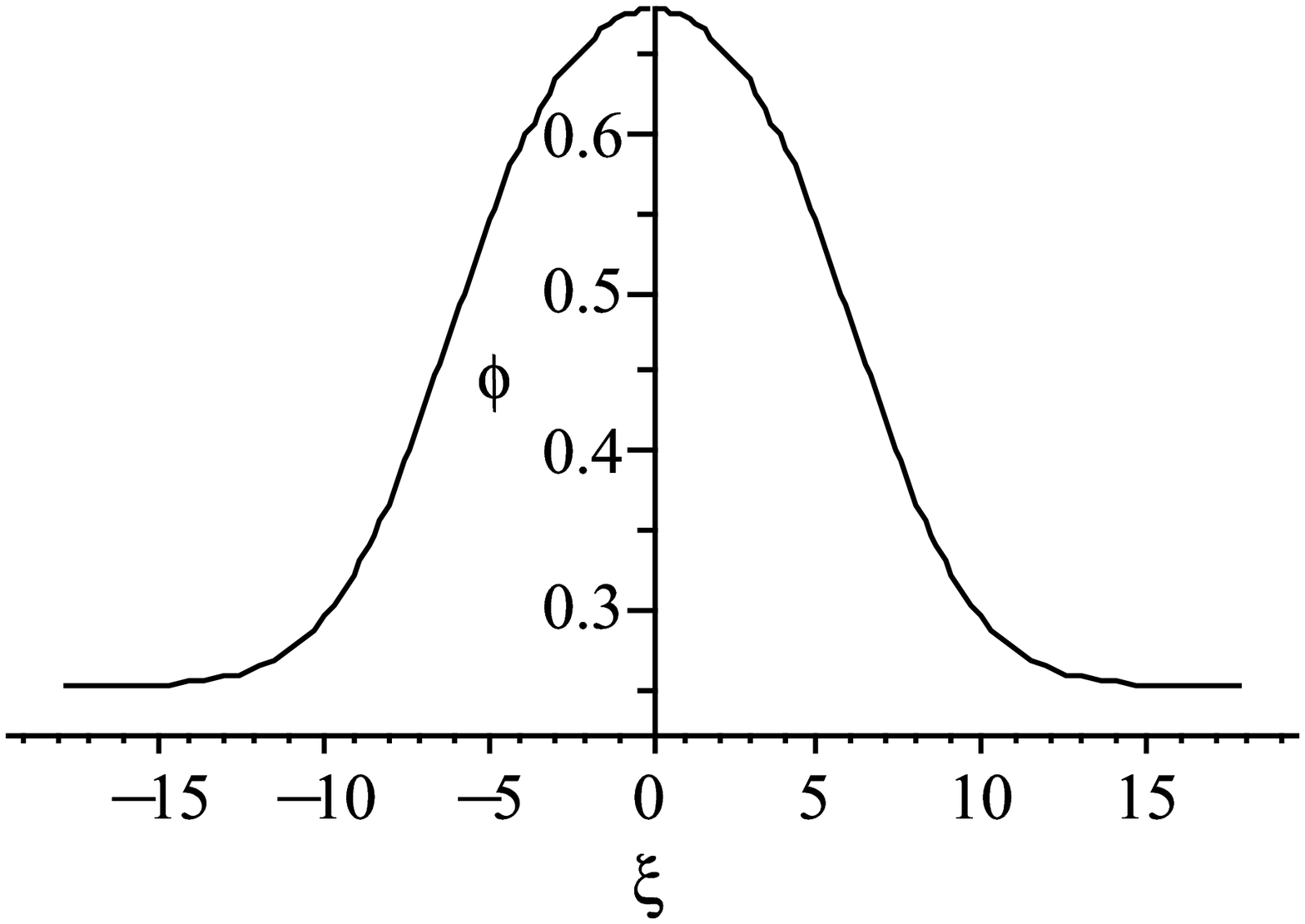}}\hspace{0.08\linewidth}
\subfloat[]{\includegraphics[height=1.5in,width=2.4in]{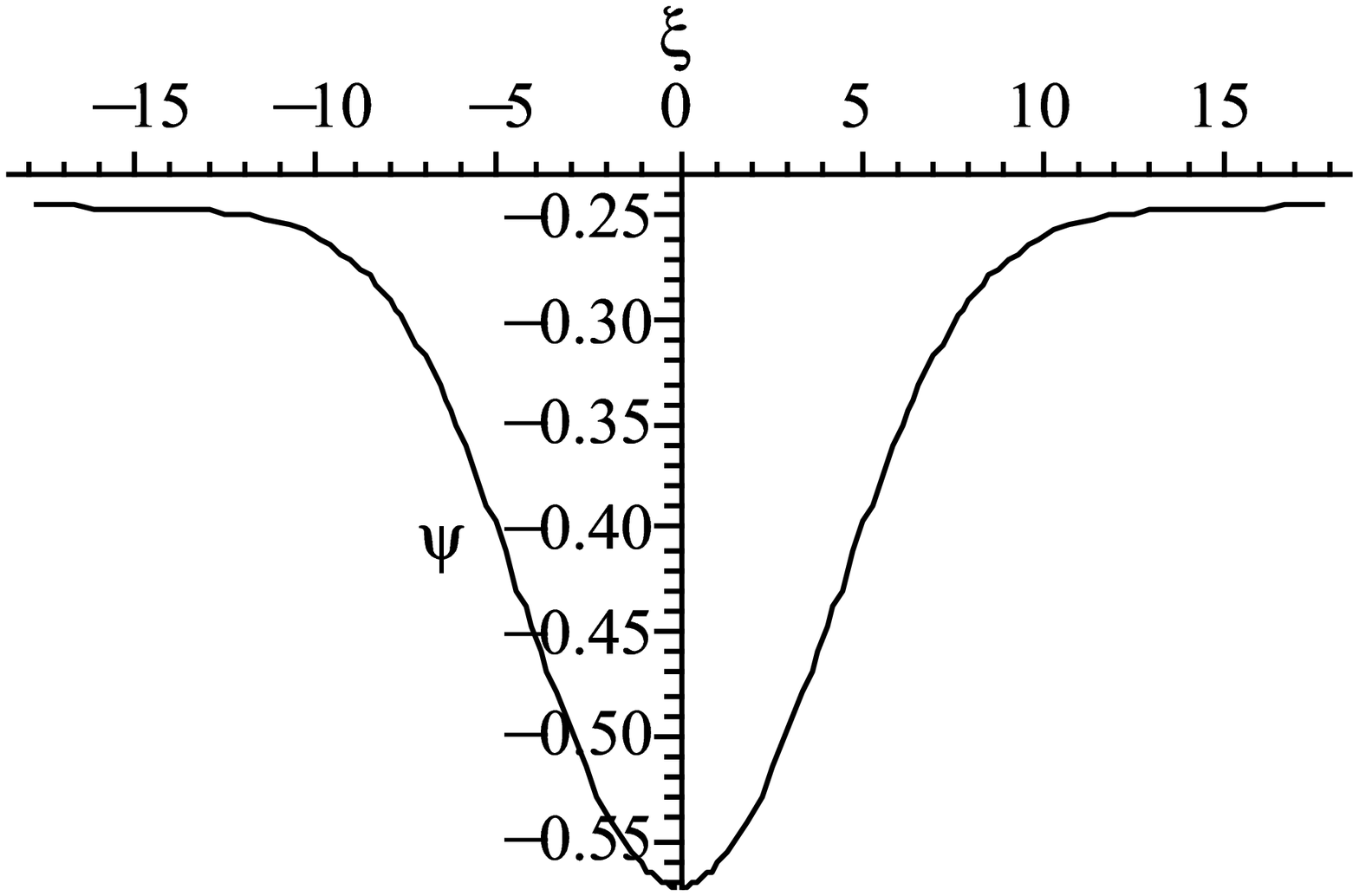}}\\
\caption{The soliton of Eqs.(\ref{eq1.2}). ($c=1$, $g=0.37$,
$h=-0.183750$)}\label{fig3}
\end{figure}

\begin{figure}[h]
\centering
\subfloat[]{\includegraphics[height=1.5in,width=2.4in]{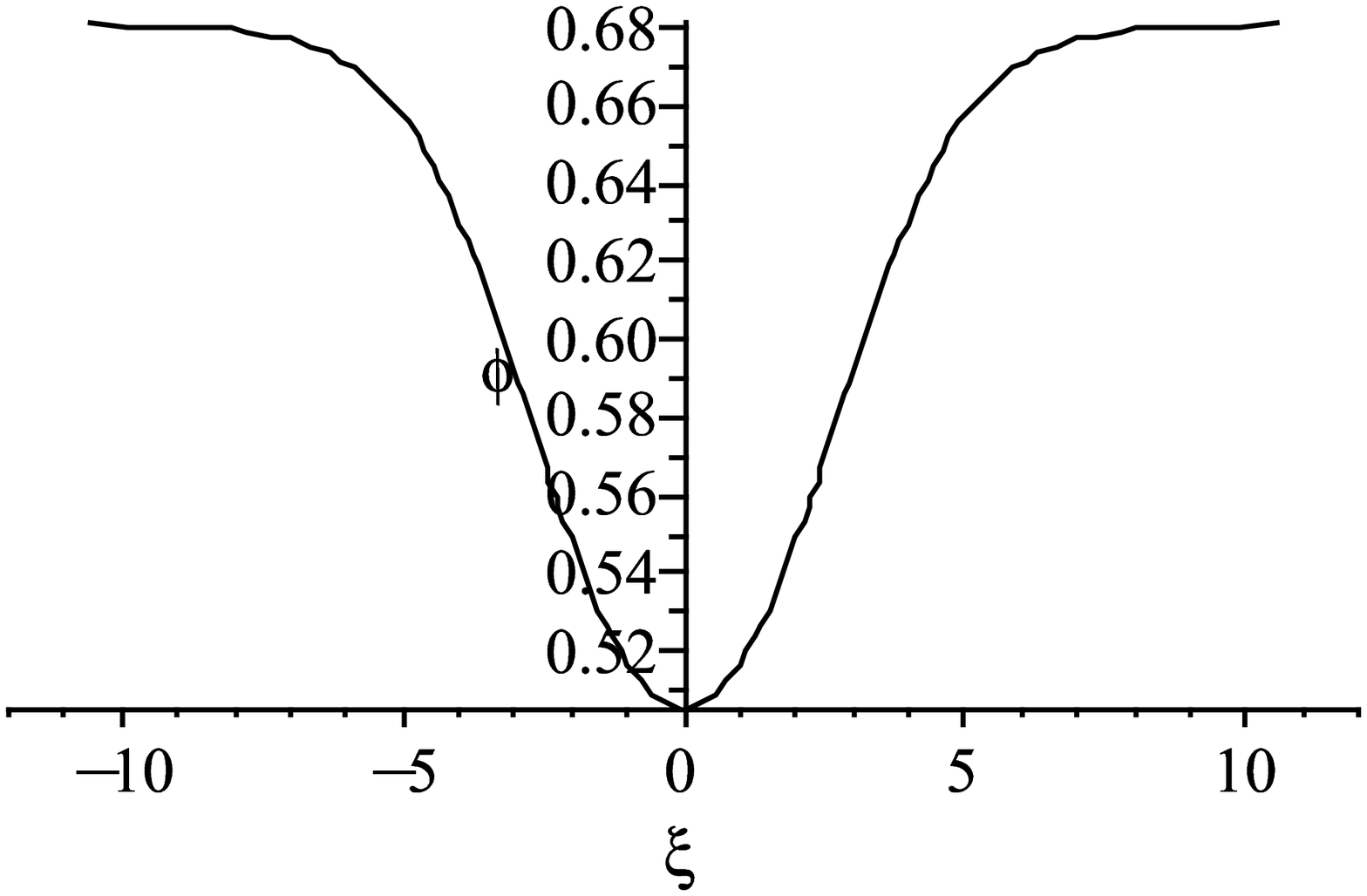}}\hspace{0.08\linewidth}
\subfloat[]{\includegraphics[height=1.5in,width=2.4in]{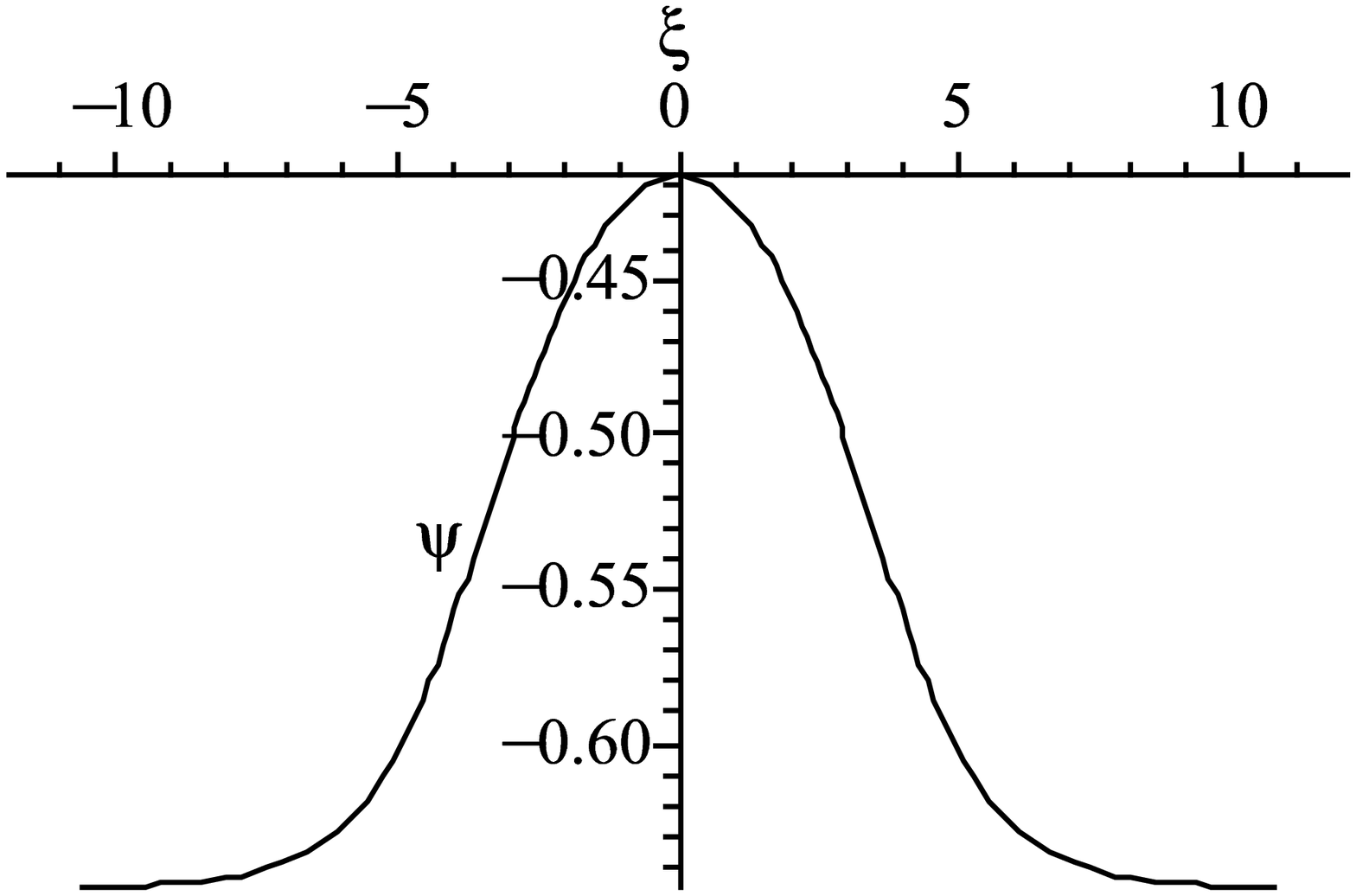}}\\
\caption{The soliton of Eqs.(\ref{eq1.2}). ($c=1$, $g=0.4$,
$h=-0.206250$)}\label{fig4}
\end{figure}

Now we take a set of data and employ Maple to display the graphs of
the above obtained soliton solutions in Figs. \ref{fig2}-\ref{fig4}.

In Fig.\ref{fig1}(e),  the heteroclinic orbits of system
(\ref{eq2.6}) or (\ref{eq2.8}) can be expressed as
\begin{equation}
\label{eq3.29}  y = \pm\frac{(\varphi- \frac{c}{4})(
\varphi-\frac{3c}{4}) }{\varphi-c} \quad \textrm{for} \quad
\frac{c}{4}\leq\varphi \leq \frac{3c}{4},
\end{equation}

Substituting Eq.(\ref{eq3.29}) into the first equation of system
(\ref{eq2.6}), and integrating along the heteroclinic orbits, we
have
\begin{equation}
\label{eq3.30}
 (\varphi_\pm- \frac{c}{4})^{\frac{3}{2}}= (\frac{3c}{4}-\varphi_\pm
 )^{\frac{1}{2}}\exp(\pm \xi)\quad \mathrm{for} \quad
\frac{c}{4}\leq\varphi_\pm \leq \frac{3c}{4},
 \end{equation}
where $\varphi_+=\varphi_+(\xi)$ is a kink solution, and
$\varphi_-=\varphi_-(\xi)$ is an antikink solution.

The corresponding solutions of the second component $v$ in
Eqs.(\ref{eq1.2}) can be expressed in parametric form as follows.
\begin{equation}
\label{eq3.31} \left\{ {\begin{array}{l}
  \xi=\pm (\frac{3}{2}\ln(\varphi_\pm- \frac{c}{4})-\frac{1}{2}\ln(\frac{3c}{4}-\varphi_\pm
 )), \\
 \psi_\pm=\frac{\textstyle c^3}{\textstyle 16(\varphi_\pm-c)},
\\
 \end{array}} \right.      \;\quad (\frac{c}{4}\leq\varphi_\pm \leq \frac{3c}{4}),
\end{equation}
with $\varphi_\pm$ as the parameter. Here $\psi_+=\psi_+(\xi)$ is an
antikink solution, and $\psi_-=\psi_-(\xi)$ is a kink solution.

\begin{figure}[h]
\centering
\subfloat[]{\includegraphics[height=1.5in,width=2.4in]{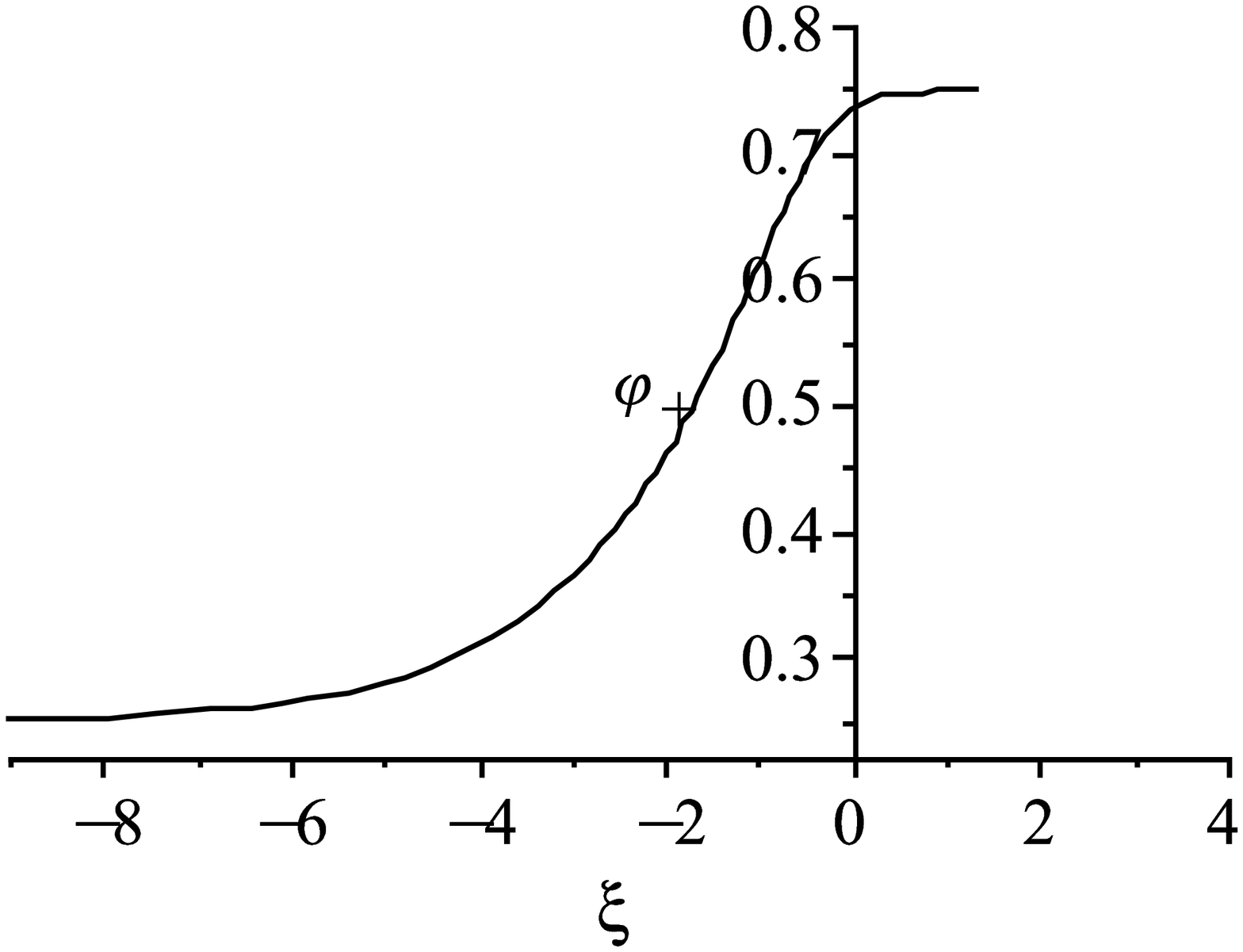}}\hspace{0.08\linewidth}
\subfloat[]{\includegraphics[height=1.5in,width=2.4in]{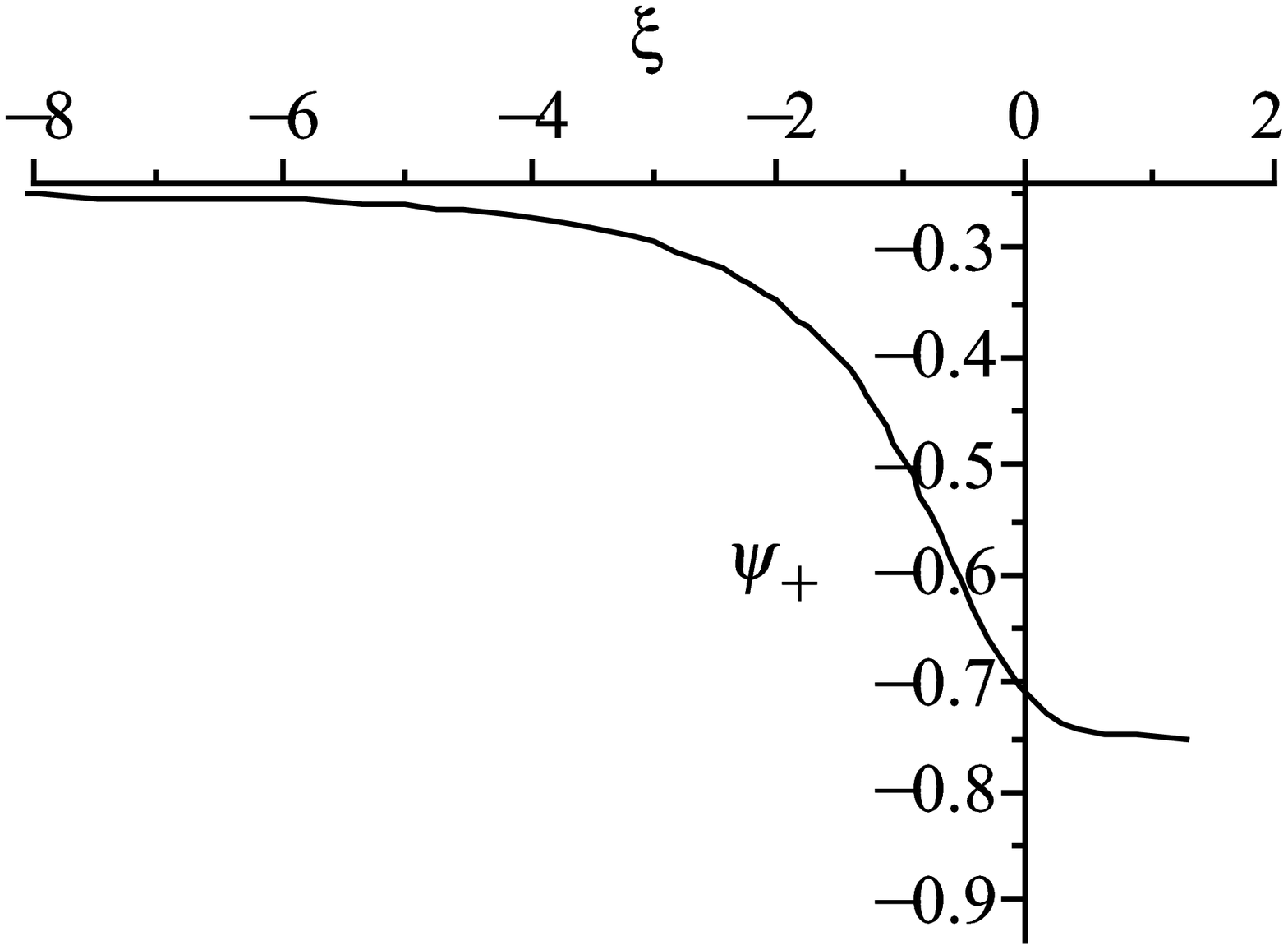}}\\
\subfloat[]{\includegraphics[height=1.5in,width=2.4in]{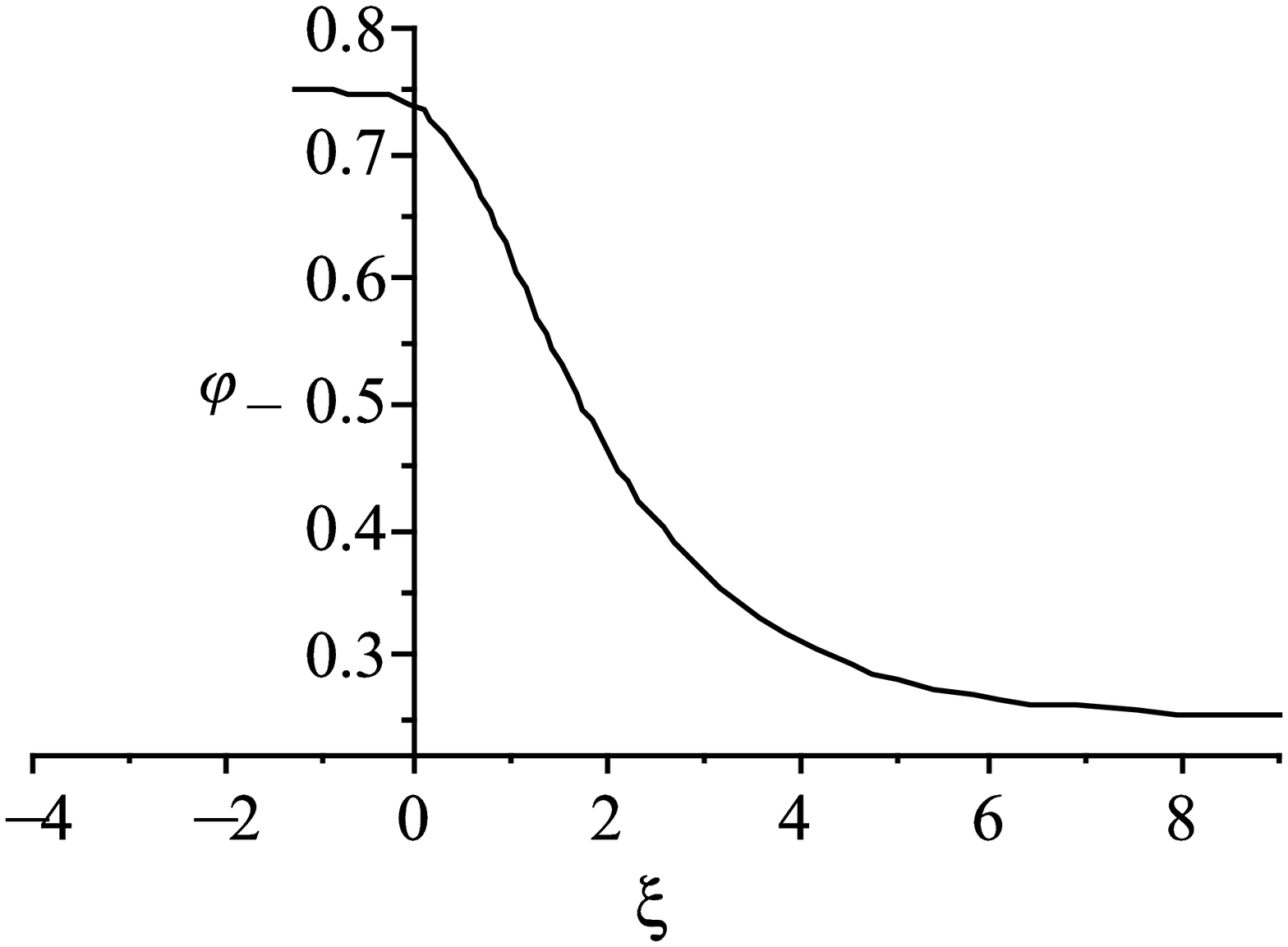}}\hspace{0.08\linewidth}
\subfloat[]{\includegraphics[height=1.5in,width=2.4in]{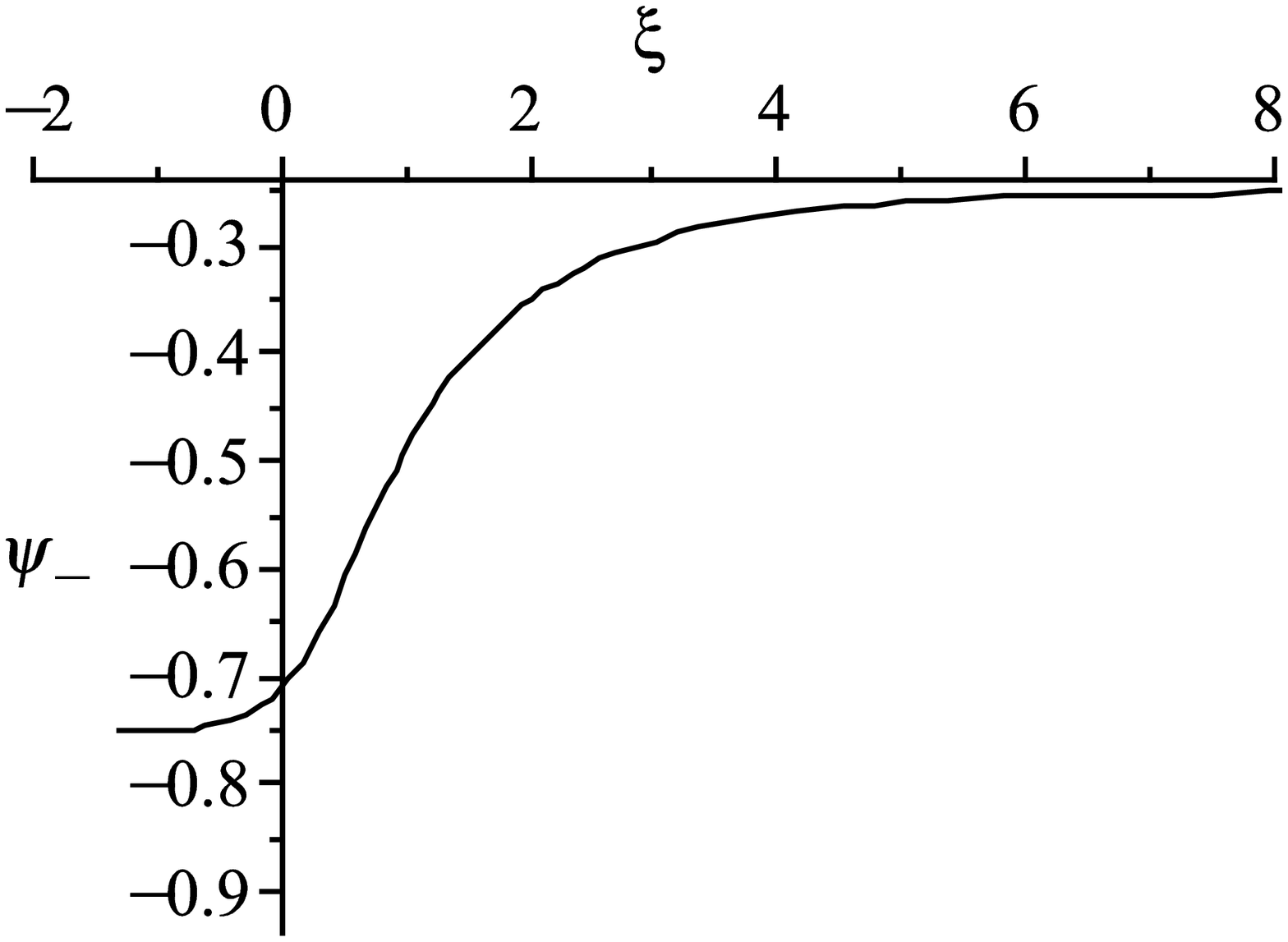}}\\
\caption{The kink and antikink wave solutions of Eqs.(\ref{eq1.2}).
($c=1$, $g=0.375$, $h=0.1875$)}\label{fig5}
\end{figure}

Also, we take a set of data and employ Maple to display the graphs
of the above obtained kink and antikink solutions in Fig.
\ref{fig5}.

\begin{remark}
(1) The Degasperis-Procesi equation (\ref{eq1.1}) does not have kink
and antikink solutions because there is no heteroclinic orbit in the
corresponding phase portraits (see \cite{23}).

(2) In \cite{24}, Vakhnenko and Parkes also suggested a generalized
Degasperis-Procesi equation. Using direct integration, they obtain
its kink and antikink solutions. So the connection between our
suggested Eqs.(\ref{eq1.2}) and the generalized Degasperis-Procesi
equation considered in \cite{24} needs further studying.

(3) The bifurcation method only enables us to analyze the restricted
class of solutions of Eqs.(\ref{eq1.2}), namely, the one-valued
solutions. However the Degasperis-Procesi equation has intriguing
solutions such as loop-like solitary wave solutions (see \cite{14}).
So whether the 2-component of Degasperis-Procesi equation
(\ref{eq1.2}) has loop-like solutions needs other mathematical
methods.

\end{remark}

\section{Conclusion}
 \label{}
 \setcounter{equation}{0}
In this work, we propose a generalization of the Degasperis-Procesi
equation, that is, a 2-component of the Degasperis-Procesi equation
(\ref{eq1.2}) for the first time. Employing the bifurcation method,
we obtain the analytic expressions for smooth soliton, kink and
antikink solutions of the 2-component of the Degasperis-Procesi
equation (\ref{eq1.2}). However, its integrability needs further
investigation.

\end{document}